\documentclass[prd,aps,nofootinbib,floatfix,11pt]{revtex4}

\usepackage{amsmath,graphicx,epsfig,amssymb,dsfont,mathtools}
\usepackage[usenames]{color}
\usepackage{ulem} %% for strike-through
\usepackage{bigstrut}
\usepackage{slashed}
\usepackage{multirow}
\usepackage{subfigure}

\allowdisplaybreaks

\begin{document}
	
\title{Analysis of the interpolating currents of baryons: anomalous dimension }
	
\author{
Zhen-Xing Zhao$^{1}$~\footnote{Email:zhaozx19@imu.edu.cn},
Yu-Ji Shi$^{2}$~\footnote{Email:shiyuji92@126.com},
Zhi-Peng Xing$^{3}$~\footnote{Email:zpxing@sjtu.edu.cn}
}

\affiliation{
$^{1}$ School of Physical Science and Technology, \\
Inner Mongolia University, Hohhot 010021, China \\
$^{2}$ School of Physics, East China University of Science and Technology, Shanghai 200237, China \\
$^{3}$ Tsung-Dao Lee Institute,
Shanghai Jiao Tong University, Shanghai 200240, China
}
		
\begin{abstract}
The anomalous dimensions for the interpolating currents
of baryons are indispensable inputs in a serious QCD analysis of baryon. 
However, the results in the literature are vague. In view of this, in this work, we investigate the one-loop anomalous dimensions for some interpolating currents such as those of $\Lambda_{Q}$
and proton. Under the requirement of precise testing the Standard Model, this study  is essential and significant for the QCD analysis in the future.
\end{abstract}

\maketitle

\section{Introduction}
Flavor physics which contains the studies of hadrons is an essential parts of particle physics and plays an important role in precise testing the Standard Model(SM) and studies Beyond Standard Model(BSM). The importance of research with hadron not only reflected by various experiment including hadron collider~\cite{BaBar:2014omp,LHCb:2018roe,Belle-II:2018jsg,BESIII:2020nme}, Cosmic Microwave Background(CMB) and Big Bang Nucleosynthesis(BBN)~\cite{Cyburt:2015mya,Planck:2019nip,ParticleDataGroup:2022pth} but also various theoretical studies~\cite{Li:2018lxi,Elor:2018twp,HeavyFlavorAveragingGroup:2022wzx,Alonso-Alvarez:2021qfd}. Recently, more and more exotic phenomenon and structures are observed in experiment~\cite{Aaij:2018dso,Belle:2019oag,Belle:2021crz,LHCb:2023zxo} and these exotic results also request the more precise predictions in theoretical studies such as perturbative QCD~\cite{Shen:2020hfq,HeavyFlavorAveragingGroup:2022wzx,Cui:2022zwm,Tao:2022hos,Tao:2023mtw}, QCD sum rules(QCDSR)~\cite{Lenz:2014jha,Zhao:2021lzd,Xin:2022qnv,Xiong:2022uwj,Aliev:2023xwr} and Lattice QCD(LQCD)~\cite{Hua:2022wop,Zhang:2021oja,Christ:2023lcc}.  

In these theoretical method we mentioned, a prevalent and necessary part of the calculation is the interpolating currents which reflects to some extent the various properties of hadrons composed of quarks. As the important non-perturbative input in exclusive perturbative QCD calculation, Light-Cone-Distribution-Amplitude(LCDA) typically constructed by the matrix element of the non-local interpolating current of hadrons~\cite{Ali:2012pn,Bali:2015ykx}.  The hadron interpolating currents are also used to build the correlation functions which are the footing stone of the QCDSR and LQCD~\cite{Sun:2023noo,Liu:2023feb}. Therefore the study of the hadron interpolating currents is essential and significant for the QCD analysis in the future.

In our previous work of QCDSR~\cite{Xing:2021enr}, we perform a careful analysis on these hadronic matrix elements with QCDSR and include the leading logarithmic(LL) correction which rely on the important property of hadron interpolating currents: anomalous dimension.  For explaining the significance of the LL correction and role of anomalous dimension in it. Let us give a brief introduction of QCDSR.

The QCDSR is a QCD-based approach to investigate
the properties of hadrons. It connects the hadron phenomenology and
QCD vacuum structure via a few universal parameters like quark condensates
and gluon condensates. It is usually thought that about 30\% uncertainty will be introduced
in QCDSR. However, as can be seen in our recent works \cite{Zhao:2020mod,Zhao:2021sje},
once a stable Borel region is found, the prediction of physical quantity
can reach high precision.
%Recently, LHCb updated a measurement of the lifetime of $\Omega_{c}$
%with $\tau(\Omega_{c}^{0})=268\pm24\pm10\pm2$ fs \cite{Aaij:2018dso},
%which is nearly four times larger than the world-average $\tau(\Omega_{c}^{0})=69\pm12$
%fs \cite{Tanabashi:2018oca} in PDG2018. Theoretically, the lifetimes
%of weak-decay baryons can be explained by the framework of heavy quark
%expansion (HQE) (see \cite{Lenz:2014jha} for a review). Among others, the calculation
%of hadron matrix elements is an important part to decipher this puzzle.
%In this direction, some efforts has been made in our recent work \cite{Zhao:2021lzd}.
In QCDSR, the Wilson coefficients of OPE should be evolved to the
energy scale $\mu_{0}$ of low-energy limit \cite{Ioffe:1981kw}. For example,
for the two-point correlation function of baryon, 
\begin{equation}
T\{J(x)\bar{J}(0)\}=\sum_{n}C_{n}(x^{2})O_{n}(0),
\end{equation}
the coefficient functions $C_{n}(p^{2})$ in the momentum space should
contain the factor 
\[
\left[\frac{\alpha_{s}(p^{2})}{\alpha_{s}(\mu_{0}^{2})}\right]^{2\gamma_{J}-\gamma_{n}},
\]
where $\gamma_{J}$ and $\gamma_{n}$ are respectively the anomalous
dimensions of the current $J$ and the operator $O_{n}$.

It can be seen that, the results of the anomalous dimensions are indispensable
inputs in a serious analysis of baryon QCD sum rules, especially for the
case of bottom baryons. However, the results in the literature are vague. In view of this, in this work, we
calculate the anomalous dimensions for some interpolating currents
such as those of $\Lambda_{Q}$ and proton.

\begin{figure}[!]
\includegraphics[width=0.6\columnwidth]{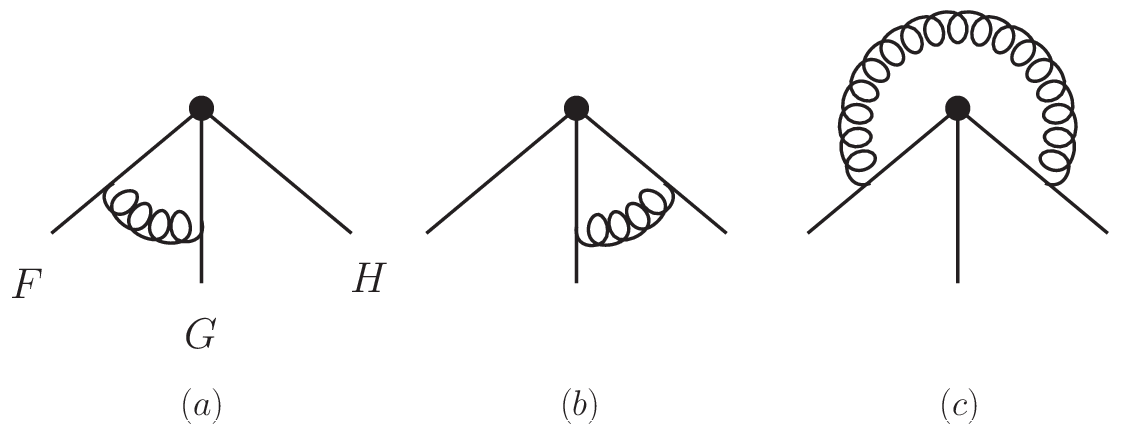}
\caption{The one-loop diagrams for the calculation of anomalous dimension.
Here $F$, $G$ and $H$ respectively stand for the quark fields of
$u$, $d$ and $Q$ for $\Lambda_{Q}$.}
\label{fig:ano_dim} 
\end{figure}

Considering the NLO, there are 3 relevant diagrams, as can be seen in Fig. \ref{fig:ano_dim}.
Following the convention of Peskin in Ref.~\cite{Peskin:1979mn}, the sum of
the divergent parts of Fig. 1a-1c will be of the form 
\begin{equation}
-\frac{4}{3}b\frac{g^{2}}{(4\pi)^{2}}\frac{\Gamma(2-d/2)}{(M^{2})^{2-d/2}}\times J,\label{eq:Peskin_convention}
\end{equation}
with $b$ some constant. Then 
\begin{align}
\gamma(g^{2}) & =\frac{4}{3}(3+2b)\frac{g^{2}}{(4\pi)^{2}}+O(g^{4}),\nonumber \\
\gamma_{J} & =\frac{2}{3}(3+2b)/\beta_{0},\label{eq:gamma_Peskin}
\end{align}
where $\beta_{0}=11-(2/3)n_{f}$ with $n_{f}$ the number of fermion
flavors. In Eq. (\ref{eq:gamma_Peskin}), the factor of $(4/3)\cdot3$
arises from the wavefunction renormalization of three fermion fields. 

The detail of deduction will be given in the rest of this paper which  is arranged as follows. In Sec. II and III,
we will calculate the anomalous dimensions for the interpolating currents
of $\Lambda_{Q}$ and proton, respectively. We conclude our paper
in the last section.

%%%%%%%%%%%%%%%%%%%%%%
%%%%%%%%%%%%%%%%%%%%%%

\section{The anomalous dimension for $\Lambda_{Q}$}
In the exclusive QCD analysis of $\Lambda_Q$, the interpolating current of it is a basic building block which is usually taken as:
\begin{equation}
J_{\Lambda_{Q}}=\epsilon_{abc}(u_{a}^{T}C\gamma_{5}d_{b})Q_{c},\label{eq:current_LQ}
\end{equation}
where $Q$ denotes a bottom or charm quark, $a,b,c$ are the color
indices and $C$ is the charge conjugate matrix. In Eq. (\ref{eq:current_LQ}),
the system of $u$ and $d$ quarks has spin and parity of $0^{+}$,
while the whole baryon has $J^{P}=1/2^{+}$.

For Fig. 1a, it can be shown that the amplitude can be calculated as a convolution of the $\Lambda_Q$ interpolating current which is seen as a trivial tree level results. The one loop results is
\begin{equation}
i{\mathcal M}_a=(-\frac{8}{3})\times ig^{2}\int\frac{d^{4}k}{(2\pi)^{4}}\frac{1}{k^{4}}\times J_{\Lambda_{Q}}.
\end{equation}
For Figs. 1b, a new structure emerges since we can apply a Fierz transformation on it as
\begin{equation}
(\bar{\psi}_{1}\gamma^{\mu}\gamma^{\nu}\psi_{2})\gamma_{\mu}\gamma_{\nu}\psi_{3}=4\ (\bar{\psi}_{1}\psi_{2})\psi_{3}-(\bar{\psi}_{1}\sigma^{\mu\nu}\psi_{2})\sigma_{\mu\nu}\psi_{3}.
\end{equation}
It turns out that 
\begin{equation}
i{\mathcal M}_b=-\frac{1}{6}\times ig^{2}\int\frac{d^{4}k}{(2\pi)^{4}}\frac{1}{k^{4}}\times\epsilon_{abc}[4\ (u_{a}^{T}C\gamma_{5}d_{b})Q_{c}-(u_{a}^{T}C\gamma_{5}\sigma^{\mu\nu}d_{b})\sigma_{\mu\nu}Q_{c}].
\end{equation}
Similarly, 
\begin{equation}
i{\mathcal M}_c=-\frac{1}{6}\times ig^{2}\int\frac{d^{4}k}{(2\pi)^{4}}\frac{1}{k^{4}}\times\epsilon_{abc}[4\ (d_{b}^{T}C\gamma_{5}u_{a})Q_{c}-(d_{b}^{T}C\gamma_{5}\sigma^{\mu\nu}u_{a})\sigma_{\mu\nu}Q_{c}].
\end{equation}
It seems that the amplitude of Fig. 1b-1c can not show the same structure with the tree level. However, fortunately, the strange structure disappear when we sum the ${\mathcal M}_b$ and ${\mathcal M}_c$ as
\begin{equation}
i{\mathcal M}_b+i{\mathcal M}_c=2\times(-\frac{2}{3})\times ig^{2}\int\frac{d^{4}k}{(2\pi)^{4}}\frac{1}{k^{4}}\times\epsilon_{abc}(u_{a}^{T}C\gamma_{5}d_{b})Q_{c}.
\end{equation}
In the above, we have used 
\begin{align}
\epsilon_{abc}(u_{a}^{T}C\gamma_{5}d_{b})Q_{c} & =\epsilon_{abc}(d_{b}^{T}C\gamma_{5}u_{a})Q_{c},\\
\epsilon_{abc}(d_{b}^{T}C\gamma_{5}\sigma^{\mu\nu}u_{a})\sigma_{\mu\nu}Q_{c} & =-\epsilon_{abc}(u_{a}^{T}C\gamma_{5}\sigma^{\mu\nu}d_{b})\sigma_{\mu\nu}Q_{c},
\end{align}
where we apply the transpose of the current of two light quarks.
Finally, the counterterm for the current is obtained as 
\begin{align}
-\delta_{J_{\Lambda_{Q}}}=i{\mathcal M}_a+i{\mathcal M}_b+i{\mathcal M}_c& =-4\times ig^{2}\int\frac{d^{4}k}{(2\pi)^{4}}\frac{1}{k^{4}}\times J_{\Lambda_{Q}} =4\times\frac{g^{2}}{(4\pi)^{2}}\frac{\Gamma(2-d/2)}{(M^{2})^{2-d/2}}\times J_{\Lambda_{Q}},
\end{align}
where $M$ is the reference scale of renormalization. As can be seen
in Eq. (\ref{eq:Peskin_convention}), $b=-3$. Therefore, the anomalous
dimension for the current of $\Lambda_{Q}$ is $-2/\beta_{0}$. It should be noticed that our result is different from the previous calculation in Ref.~\cite{Ovchinnikov:1991mu}. This different also indicate the important of our work.

\section{The anomalous dimension for proton}
Proton as the one of most popular hadron of the universe play a very important role in particle physics studies. The interpolating current for proton is usually taken as \cite{Ioffe:1981kw}:
\begin{equation}
J_{p}=\epsilon_{abc}(u_{a}^{T}C\gamma^{\mu}u_{b})\gamma_{5}\gamma_{\mu}d_{c},
\end{equation}
where the system of two $u$ quarks has spin and parity of $1^{+}$,
while the whole baryon has $J^{P}=1/2^{+}$.

Without loss of generality, in this section, we assume that the current
is built from quark fields of three distinct flavors $F$, $G$, $H$:
\begin{equation}
J_{p}=\epsilon_{abc}(F_{a}^{T}C\gamma^{\mu}G_{b})\gamma_{5}\gamma_{\mu}H_{c}.\label{eq:current_proton}
\end{equation}
According to the Fig. 1a, amplitude can be shown that 
\begin{equation}
i{\mathcal M}_a=(-\frac{2}{3})\times ig^{2}\int\frac{d^{4}k}{(2\pi)^{4}}\frac{1}{k^{4}}\times J_{p}.
\end{equation}
For Fig. 1b, a new structure emerges because the Fierz transformation as
\begin{eqnarray}
 &  & (\bar{\psi}_{1}\gamma^{\mu}\gamma^{\nu}\gamma^{\rho}\psi_{2})\gamma_{5}\gamma_{\mu}\gamma_{\nu}\gamma_{\rho}\psi_{3}\nonumber \\
 & = & (\bar{\psi}_{1}\gamma^{\mu}\gamma^{\nu}\gamma^{\rho}\psi_{2})\gamma_{5}(2g_{\nu\rho}\gamma_{\mu}-2g_{\mu\rho}\gamma_{\nu}+\gamma_{\rho}\gamma_{\mu}\gamma_{\nu})\psi_{3}\nonumber \\
 & = & 10\ (\bar{\psi}_{1}\gamma^{\mu}\psi_{2})\gamma_{5}\gamma_{\mu}\psi_{3}-6\ (\bar{\psi}_{1}\gamma^{\mu}\gamma_{5}\psi_{2})\gamma_{\mu}\psi_{3}.
\end{eqnarray}
It turns out that 
\begin{equation}
i{\mathcal M}_b=-\frac{1}{6}\times ig^{2}\int\frac{d^{4}k}{(2\pi)^{4}}\frac{1}{k^{4}}\times\epsilon_{abc}[10\ (F_{a}^{T}C\gamma^{\mu}G_{b})\gamma_{5}\gamma_{\mu}H_{c}-6\ (F_{a}^{T}C\gamma^{\mu}\gamma_{5}G_{b})\gamma_{\mu}H_{c}].
\end{equation}
Similarly, 
\begin{equation}
i{\mathcal M}_c=-\frac{1}{6}\times ig^{2}\int\frac{d^{4}k}{(2\pi)^{4}}\frac{1}{k^{4}}\times\epsilon_{abc}[10\ (G_{a}^{T}C\gamma^{\mu}F_{b})\gamma_{5}\gamma_{\mu}H_{c}-6\ (G_{a}^{T}C\gamma^{\mu}\gamma_{5}F_{b})\gamma_{\mu}H_{c}].
\end{equation}
Similar to the calculation of $\Lambda_Q$ interpolating current we can combine the ${\mathcal M}_b$ and ${\mathcal M}_c$ together as
\begin{equation}
i{\mathcal M}_b+i{\mathcal M}_c=2\times(-\frac{5}{3})\times ig^{2}\int\frac{d^{4}k}{(2\pi)^{4}}\frac{1}{k^{4}}\times\epsilon_{abc}(F_{a}^{T}C\gamma^{\mu}G_{b})\gamma_{5}\gamma_{\mu}H_{c}.
\end{equation}
In the above, we have used 
\begin{align}
\epsilon_{abc}(G_{a}^{T}C\gamma^{\mu}F_{b})\gamma_{5}\gamma_{\mu}H_{c} & =\epsilon_{abc}(F_{a}^{T}C\gamma^{\mu}G_{b})\gamma_{5}\gamma_{\mu}H_{c},\\
\epsilon_{abc}(G_{a}^{T}C\gamma^{\mu}\gamma_{5}F_{b})\gamma_{\mu}H_{c} & =-\epsilon_{abc}(F_{a}^{T}C\gamma^{\mu}\gamma_{5}G_{b})\gamma_{\mu}H_{c}.
\end{align}
Finally, the counterterm for the current is obtained as 
\begin{align}
-\delta_{J_{p}}=\text{Fig. 1a}+\text{Fig. 1b}+\text{Fig. 1c} & =-4\times ig^{2}\int\frac{d^{4}k}{(2\pi)^{4}}\frac{1}{k^{4}}\times J_{p} =4\times\frac{g^{2}}{(4\pi)^{2}}\frac{\Gamma(2-d/2)}{(M^{2})^{2-d/2}}\times J_{p}.
\end{align}
It happens that $\delta_{J_{p}}=\delta_{J_{\Lambda_{Q}}}$ and thereby
$\gamma_{J_{p}}=\gamma_{J_{\Lambda_{Q}}}=-2/\beta_{0}$. It should be noticed that, our result
is consistent with previous calculation in Ref.~\cite{Peskin:1979mn} and the correctness of our work are confirmed.

\section{Conclusions}

The anomalous dimensions for the interpolating currents
of baryons are indispensable inputs in a serious analysis of baryon exclusive QCD analysises. However, the results in the literature are vague. Therefore in our work we investigate the one-loop anomalous dimensions for some interpolating currents such as those of $\Lambda_{Q}$
and proton. Since the interpolating currents play an very important and fundamental role in QCD analysis such as perturbative QCD, QCD sum rules(QCDSR) and Lattice QCD(LQCD), our work will be very helpful for precise QCD studies in the future.

In this work, we do not consider the interpolating current of $\Delta$,
which has the quark content of $uuu$ and spin-3/2. We feel that the
interpolating current of $\Delta$ written in terms of Dirac gamma
matrices
\begin{equation}
(J_{\Delta})^{\mu}=\epsilon_{abc}(u_{a}^{T}C\gamma^{\mu}u_{b})u_{c},
\end{equation}
is cumbersome so that it is not easy to see clearly the completely
symmetrical relation among these three $u$ quarks. It seems that
the presription in \cite{Peskin:1979mn} is still a better choice and the analysis of the $\Delta$ interpolating current  will going on in the future studies.

\section*{Acknowledgements}

The authors are grateful to Prof. Zhi-Gang Wang for valuable discussions
and to Prof. Wei Wang for constant encouragements. This work is supported
in part by National Natural Science Foundation of China under Grant
No. 12065020, by China Postdoctoral Science Foundation under Grant No. 2022M72210.

\end{document}